# Iodine versus Bromine Functionalization for Bottom-Up Graphene Nanoribbon Growth: Role of Diffusion

Christopher Bronner*,#,1, Tomas Marangoni#,2, Daniel J. Rizzo[1], Rebecca Durr[2], Jakob Holm Jørgensen[3], Felix R. Fischer*,2,4,5 and Michael F. Crommie*,1,4,5

[1]Department of Physics, University of California, Berkeley, CA 94720, United States

[2]Department of Chemistry, University of California, Berkeley, CA 94720, United States

[3]iNANO and Department of Physics and Astronomy, Aarhus University, 8000 Aarhus, Denmark

[4]Materials Sciences Division, Lawrence Berkeley National Laboratory, Berkeley, CA 94720, United States

[5]Kavli Energy NanoSciences Institute at the University of California Berkeley and the Lawrence Berkeley National Laboratory, Berkeley, California 94720, United States

* Corresponding authors (CB: +1 (510) 642-6882, bronner@berkeley.edu; FF: +1 (510) 643-7205, ffischer@berkeley.edu; MC: +1 (510) 642-9392, crommie@berkeley.edu)

# These authors contributed equally to this work.

# Abstract

Deterministic bottom-up approaches for synthesizing atomically well-defined graphene nanoribbons (GNRs) largely rely on the surface-catalyzed activation of selected labile bonds in a molecular precursor followed by step growth polymerization and cyclodehydrogenation. While the majority of successful GNR precursors rely on the homolytic cleavage of thermally labile C–Br bonds, the introduction of weaker C–I bonds provides access to monomers that can be polymerized at significantly lower temperatures, thus helping to increase the flexibility of the GNR synthesis process. Scanning tunneling microscopy (STM) imaging of molecular precursors, activated intermediates, and polymers resulting from stepwise thermal annealing of both Br and I substituted precursors for chevron GNRs reveals that the polymerization of both precursors proceeds at similar temperatures on Au(111). This observation is consistent with diffusion-limited polymerization of the surface-stabilized radical intermediates that emerge from homolytic cleavage of either the C–Br or the C–I bonds.

## 1. Introduction

Graphene nanoribbons (GNRs), nanometer-wide strips of single-layer carbon, blend many of the exotic electronic properties observed in graphene sheets with a structurally tunable band gap.[1,2,3,4] Quantum effects imposed by parameters such as width, length, edge symmetry, and doping pattern allow the GNR band structure to be rationally designed from the bottom up.[1,2,4,5,6,7] While a variety of top-down approaches have pioneered the synthesis of GNRs, the harsh reaction conditions and the limited structural control of lithography-based manufacturing techniques have thus far prevented access to nanoscale atomically precise GNRs with the sizeable bandgaps required for many electronics applications.[3,8,9,10,11] A rational bottom-up synthesis scheme,[12,13,14,15] on the other hand, has met these challenges both through traditional solution-based approaches[16,17,18,19,20] and surface-catalyzed reactions.[21,22,23,24,25,26]

The surface-assisted synthesis of GNRs relies on a two-step process. In the first stage thermally activated molecular precursors polymerize on the surface via a step-growth mechanism at T ≈ 200 °C. In the second stage the resulting polymers undergo a cyclodehydrogenation reaction T ≈ 400 °C that leads to an extended π-system across the GNR.[21,26,27,28,29,30,31,32] Temperature-programmed x-ray photoelectron spectroscopy performed during the synthesis of N=7 armchair GNRs (7AGNRs) on Au(111) has shown that the homolytic cleavage of C–Br bonds is initiated at 100 °C.[27,28] The dissociated halogen atoms remain bound to the surface[27,28] until they eventually desorb associatively as HBr during cyclodehydrogenation.[29] Besides halogen dissociation, the polymerization step involves the diffusion and subsequent recombination of the radical intermediates. For small monomers the initial dehalogenation represents the rate-limiting step, while for larger precursors diffusion becomes increasingly important.[33]

Lowering the activation/polymerization temperature for this process is desirable in order to reduce unwanted side-reactions that lead to defects and irreversible premature chain termination during polymerization. A sufficiently low dehalogenation barrier could potentially even allow bottom-up synthesis of GNRs on insulators,[34] an essential requirement for future electronic device fabrication. One strategy to reduce the dehalogenation barrier is the use of a substrate that is more catalytically active than Au(111). Indeed, on Ag(111)[35,36,37] and Cu(111)[38,39,40] the C–Br bond dissociates at temperatures as low as room temperature. However, increased catalytic activity can also induce unselective C–H dissociation[27,41,42] as well as undesired bond rearrangements.[43,44] An alternative strategy relies on the weakening of the carbon-halogen bond itself (see Fig. 1a). The C–I bond is weaker than the C–Br bond and calculations even predict a lower barrier for deiodination on Au(111) than for debromination on Ag(111).[33] Iodinated monomers have previously been activated on all three coinage metals[45,14,15,46] as well as some insulators.[34,47] Experiments show C–I bond dissociation on Au(111) at room temperature[45,48,49] and below,[50,51] a significant reduction compared to the debromination temperature on Au(111).

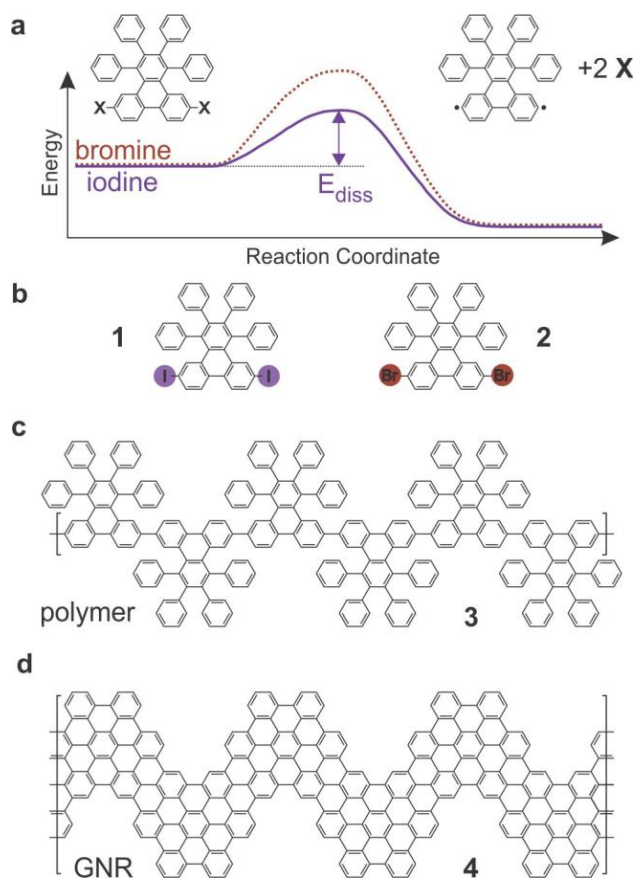

**Figure 1:** (a) Schematic representation of the reduction in bond dissociation barrier for iodinated molecular precursors compared to brominated precursors. (b) Structure of the iodinated (**1**) and brominated (**2**) molecular precursors. (c) *poly*-**1** obtained through surface-catalyzed step growth polymerization. (d) Fully cyclized chevron GNR **4**.

Here we report a comparative study of surface-supported synthesis of chevron GNRs from iodinated and brominated molecular precursors that are otherwise identical. Scanning tunneling microscopy (STM) of reaction intermediates for various annealing temperatures reveals that even though C–I bonds cleave at a significantly lower temperature than C–Br bonds, both types of halogenated monomers polymerize at the same temperature. We thus conclude that the rate-limiting step in the polymerization of **1** is the diffusion and recombination of surface-stabilized radical intermediates.

## 2. Experimental Methods

Molecular precursors **1** and **2** (Fig.1b) featuring iodide and bromide substituents respectively were prepared following literature procedures.[21,52] The Au(111) single crystal substrate was cleaned by cycles of sputtering and annealing. **1** and **2** were deposited onto the clean Au(111) surface from a home-made Knudsen cell evaporator held at 160 °C and 215 °C, respectively. The substrate temperature during evaporation was held at $T < -50$ °C (the sample was taken directly from the cryogenic STM sample stage for deposition). Samples were annealed at successively increasing temperatures for 10 min per annealing cycle and then imaged after each annealing cycle in a home-built STM kept at $T = 13$ K.

## 3. Results

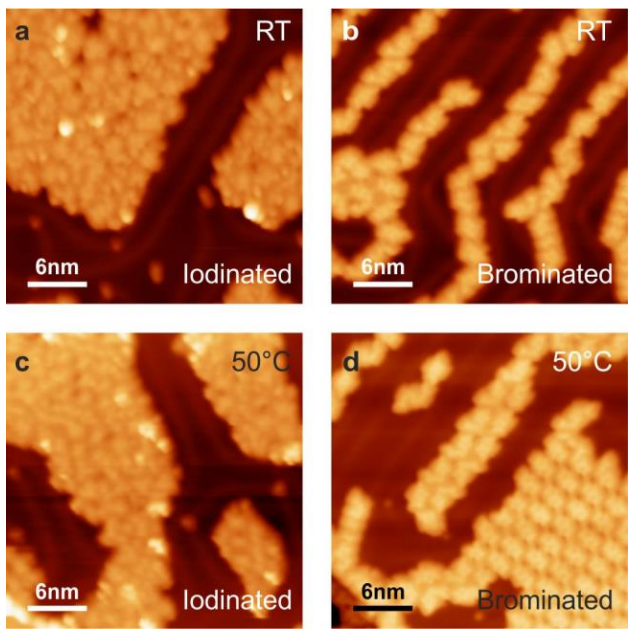

**Figure 2:** (a) **1** on Au(111) after annealing at room temperature ($V$ = 2.0V, $I$ = 20pA). (b) **2** on Au(111) after annealing at room temperature ($V$ =2.0V, $I$ = 20pA). (c) **1** on Au(111) after annealing at 50 °C ($V$ =2.0V, $I$ = 10pA). (d) **2** on Au(111) after annealing at 50 °C ($V$ =2.0V, $I$ = 20pA).

Fig. 2a shows an STM image of the Au(111) surface decorated with a sub-monolayer of iodinated precursor **1** that was allowed to equilibrate at room temperature before being cooled to 13K for imaging. For comparison, Fig. 2b shows the brominated precursor **2** prepared under the same conditions. Both species preferentially adsorb along the herringbone reconstruction and aggregate into islands as the coverage increases. A noticeable difference, however, can be seen between the islands of precursors **1** and **2**. The brominated monomers (precursor **2**) appear to be less disordered than those formed by the iodinated monomers (precursor **1**). This difference is even more striking after annealing the sample to 50 °C as shown in Figs. 2c,d. Here the level of disorder in precursor **1** islands remains essentially unchanged, whereas precursor **2** islands have become significantly more well-ordered. This suggests that precursor **1** is much less mobile on the Au(111) surface at these temperatures than precursor **2**.

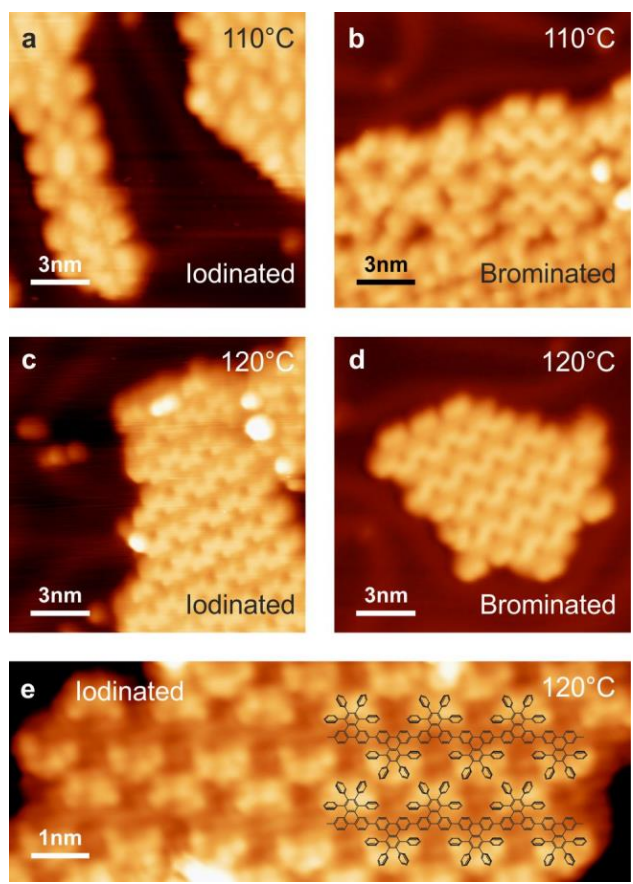

**Figure 3**: (a) Iodinated monomers **1** become more well ordered after annealing at 110 °C ($V$ =2.0V, $I$ = 20pA). (b) Brominated monomers **2** become more well ordered after annealing at 110 °C ($V$ =1.0V, $I$ = 20pA). (c) Polymer islands formed from **1** after annealing at 120 °C ($V$ = 2.0V, $I$ = 10pA). (d) Polymer islands formed from **2** after annealing at 120 °C ($V$ =1.0V, $I$ = 10pA). (e) Magnified image of polymer island obtained from **1** ($V$ =2.0V, $I$ = 40pA) with superimposed polymer structure.

This difference in behavior between the two types of precursors disappears as the temperature is raised to $T > 110$ °C. In this temperature range both precursor types undergo a gradual transition into similar highly ordered structures on the surface. This can be seen in Figs. 3a,b which show both precursor types after annealing to an intermediate state at $T = 110$ °C, but is most striking in the image taking after annealing at $T = 120$ °C, as shown in Figs. 3c,d. Here both species adopt virtually identical striped chevron island configurations. Fig. 3e shows a close-up view of such an island obtained from precursor **1** after a $T = 120$ °C anneal. The ordered phase seen here can confidently be assigned to a covalently linked polymer **3** structure that self-assembles into densely packed islands.[21] The highest points of this polymer phase have an apparent height of 3.3 Å and correspond to the four non-planar phenyl rings lining the convex edges of the polymer, while the apparent height of the conjugated planar backbone is only 2.5 Å. A periodicity of 1.6 nm between adjacent protrusions is consistent with the expected distance between covalently linked monomers. The appearance of polymer **3** formed from iodinated precursor **1** and from brominated precursor **2** is identical.[21] Regardless of which halide is incorporated in the monomer precursor and despite the substantial differences in activation barrier for the different halide bonds, the polymerization of the surface-stabilized radical intermediates proceed in the same temperature range: 50 °C $< T <$ 120 °C.

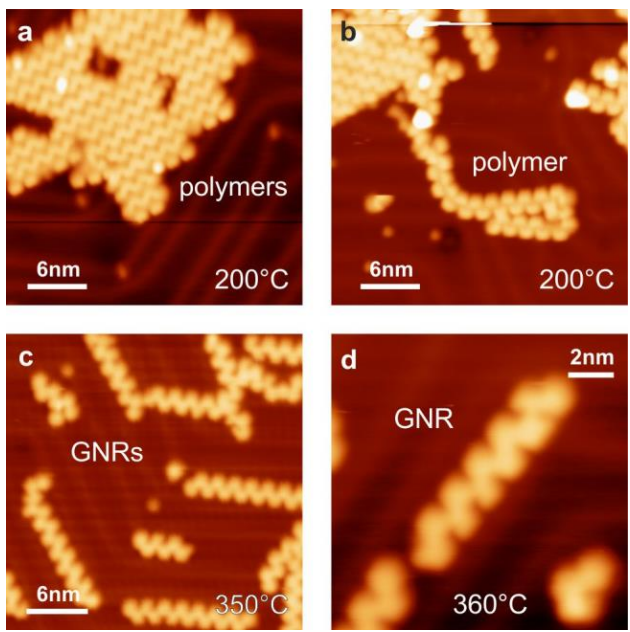

**Figure 4:** (a) Self-assembled polymers obtained from iodinated monomer **1** on Au(111) after annealing at 210 °C ($V$ = 1.0V, $I$ = 20pA). (b) Isolated polymer observed after annealing the monomers of (a) at 250 °C ($V$ = 2.0V, $I$ = 20pA). (c) Fully cyclized chevron GNRs **4** on Au(111) obtained after annealing the polymers of (b) at 350 °C ($V$ = 1.0V, $I$ = 20pA). (d) Magnified image of chevron GNR obtained after annealing the GNRs of (c) at 360 °C ($V$ = 1.0V, $I$ = 20pA).

The well-ordered structure of the polymer islands remained unchanged even when samples were heated to $T > 200$ °C, exceeding temperatures that are typically used to induce polymerization during GNR synthesis for brominated precursors (Fig. 4a). Further annealing at temperatures up to $T = 300$ °C leads to a decrease in the size of islands as individual polymer chains diffuse across the surface (Fig. 4b). The curved shape of frequently observed individual polymers (Fig. 4b) is further evidence of the covalent nature of the bond between monomers. Annealing the polymers to $T = 350$ °C induces sequential cyclodehydrogenation and leads to fully extended, flat chevron GNRs (Figs. 4c,d) with an apparent height of 2.0 Å. The absence of GNR islands is consistent with the lack of nonplanar phenyl rings (that interact through π-stacking interactions) in the GNR phase compared to the polymer phase.

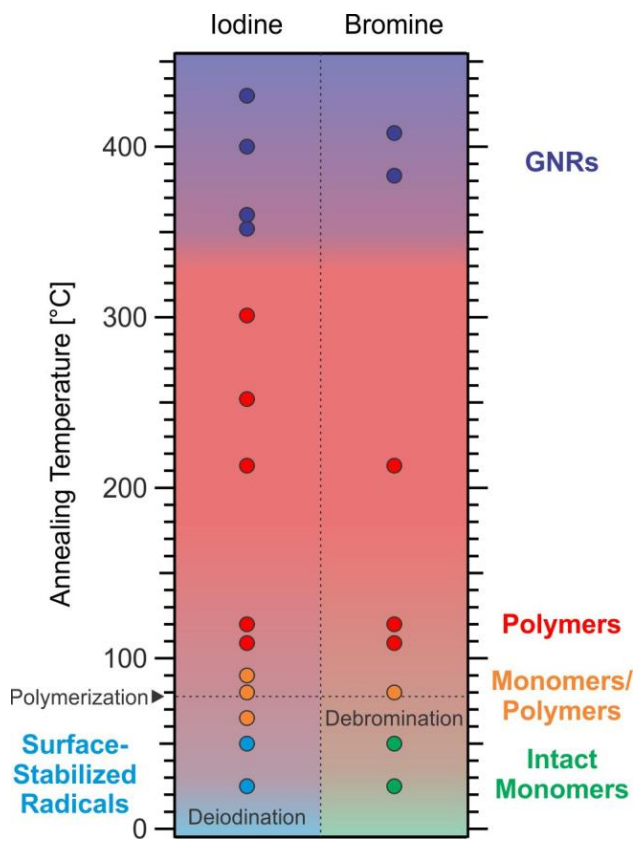

**Figure 5:** Phase diagram for synthesis of chevron GNRs from iodinated **1** and brominated **2** monomers. Color coding: blue circles indicate surface-stabilized radicals (Figs. 2a,c), green circles indicate intact monomers (Figs. 2b,d), orange circles indicate mixed phases comprised of monomers and polymers (Figs. 3a,b), red circles indicate the polymer phase (Figs. 3c,d), purple circles indicate the fully cyclized chevron GNR phase.

The full phase diagram for both the iodinated and brominated molecules can be seen in Fig. 5. Both materials show very similar polymerization and cyclodehydrogenation behavior above $T = 50$ °C. For T > 120 °C only polymers are observed in self-assembled islands while for T > 350 °C only GNRs are observed. It is in the lower temperature range that we see a marked difference in behavior, arising mainly from the reduced diffusivity of the iodinated precursors compared to the brominated ones (Fig. 2). This can be explained by room temperature dehalogenation of the iodinated precursors which results in radicals that are more tightly bound to the gold surface than halogen-passivated monomers.[45,48,49] The brominated precursors, on the other hand, remain passivated by Br in this temperature range and thus diffuse more freely on the surface. This behavior is consistent with the weaker C–I bond which is expected to cleave at lower temperature than the C–Br bond.[27,28,45,48,49] As a result, the phase diagram shows precursor **1** as surface-stabilized radicals for T < 80 °C while precursor **2** remains intact in this temperature range.

## 4. Summary and Discussion

We conclude that, contrary to our expectations based on studies of other aryl-halides on Au(111),[27,28,45,48,49] a reduction in the bond dissociation energy by switching from C–Br to C–I bonds does *not* lead to a reduction in the polymerization temperature. Even though previous experiments have demonstrated dehalogenation[45,48,49] (and even covalent C–C coupling[49,53]) for aryl-iodides on Au(111) at room temperature, the iodinated monomers **1** explored here did not polymerize at room temperature. This is explained by the reduced diffusivity of surface-stabilized radical intermediates which becomes the rate-limiting step for polymerization, even though dehalogenation occurs at lower temperatures for iodinated precursors compared to brominated precursors. This is consistent with previous studies showing that aryl radicals with increasing size on Au(111) require a continuous increase in temperature to facilitate surface diffusion and subsequent homo-coupling.[45,49,53,46,54] These results highlight the importance of understanding molecular diffusion in addition to the dissociation behavior of halogen substituents in order to perform rational on-surface synthesis.

## Acknowledgments

Research supported by the Office of Naval Research MURI Program (molecular deposition, sample preparation), by the U.S. Department of Energy (DOE), Office of Science, Basic Energy Sciences (BES), under award no. DE-SC0010409 (design, synthesis, and characterization of molecular precursors) and Nanomachine Program award no. DE-AC02-05CH11231 (surface reaction characterization), and by DARPA, the U. S. Army Research Laboratory and the U. S. Army Research Office under contract/grant number W911NF-15-1-0237 (image analysis). C.B. acknowledges support through the Fellowship Program of the German National Academy of Sciences Leopoldina under grant no. LPDS 2014-09.

## Supporting Information

Synthesis methods for the iodinated precursor and STM images of the iodinated and brominated precursors on Au(111) after annealing at various temperatures.

[39] Chen, M., Xiao, J., Steinrück, H.-P., Wang, S., Wang, W., Lin, N., Hieringer, W., and Gottfried, J. M.: Combined Photoemission and Scanning Tunneling Microscopy Study of the Surface-Assisted Ullmann Coupling Reaction. *J. Phys. Chem. C* 118, 6820-6830 (2014)

[40] Gutzler, R., Cardenas, L., Lipton-Duffin, J., El Garah, M., Dinca, L. E., Szakacs, C. E., Fu, C., Gallagher, M., Vondráček, M., Rybachuk, M., Perepichka, D. F., and Rosei, F.: Ullmann-type coupling of brominated tetrathienoanthracene on copper and silver. *Nanoscale* 6, 2660 (2014)

[41] Han, P., Akagi, K., Federici Canova, F., Mutoh, H., Shiraki, S., Iwaya, K., Weiss, P. S., Asao, N., and Hitosugi, T.: Bottom-Up Graphene-Nanoribbon Fabrication Reveals Chiral Edges and Enantioselectivity. *ACS Nano* 8, 9181-9187 (2014)

[42] Sánchez-Sánchez, C., Dienel, T., Deniz, O., Ruffieux, P., Berger, R., Feng, X., Müllen, K., and Fasel, R.: Purely Armchair or Partially Chiral: Non-contact Atomic Force Microscopy Characterization of Dibromo-Bianthryl-Based Graphene Nanoribbons Grown on Cu(111). *ACS Nano* 10, 8006-8011 (2016)

[43] Fan, Q., Wang, C., Liu, L., Han, Y., Zhao, J., Zhu, J., Kuttner, J., Hilt, G., and Gottfried, J. M.: Covalent, Organometallic, and Halogen-Bonded Nanomeshes from Tetrabromo-Terphenyl by Surface-Assisted Synthesis on Cu(111). *J. Phys. Chem. C* 118, 13018-13025 (2014)
20

**TOC Graphic**

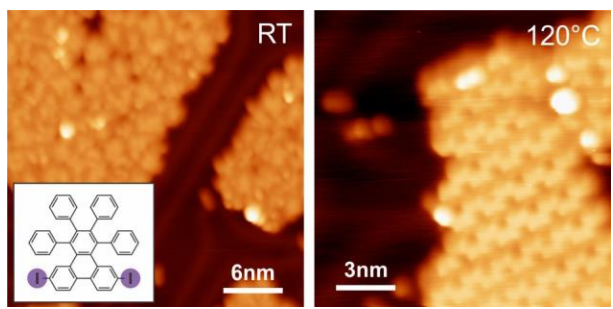